\documentclass[twocolumn, amssymb, amsmath, nobibnotes, nofootinbib, aps, prb,showpacs]{revtex4-1}
\usepackage[dvips]{graphicx}
\begin{document}
\title{Short-range magnetic correlation and magnetodielectric coupling in multiferroic Pb$_3$TeMn$_3$P$_2$O$_{14}$}
\author{Rafikul Ali Saha$^1$}
\email{msras@iacs.res.in}
%\author{Tanusri Saha-Dasgupta$^{2}$}
\author{Desheng Fu$^2$}
\author{Mitsuru Itoh$^3$}
\author{Sugata Ray$^{1}$}
%\email{mssr@iacs.res.in}
\affiliation{$^1$School of Materials Sciences, Indian Association for the Cultivation of Science, 2A \& 2B Raja S. C. Mullick Road, Jadavpur, Kolkata 700 032, India}
%\affiliation{$^2$Department of Condensed Matter Physics and Material Sciences, S. N. Bose National Centre for Basic Sciences, Block JD, Sector 3, Saltlake, Kolkata -700106, India}
\affiliation{$^2$Department of Electronics and Materials Science, and Department of Optoelectronics and Nanostructure Science, Graduate School of Science and Technology, Shizuoka University, 3-5-1 Johoku, Naka-ku, Hamamatsu 432-8561, Japan}
\affiliation{$^3$Materials and Structures Laboratory, Tokyo Institute of Technology, 4259 Nagatsuta, Yokohama 226-8503, Japan}
\pacs {}
\begin{abstract}
In this paper the structural, magnetic, and dielectric properties of langasite compound Pb$_3$TeMn$_3$P$_2$O$_{14}$ have been investigated as a candidate of short-range magnetic correlations driven development of dielectric anomaly above N$\acute{e}$el temperature of ($T_N$=) 7 K. Presence of dielectric anomaly, structural phase transition and a short range magnetic correlation at the same temperature (at around 100 K) as well as magnetic field dependent capacitance clearly indicate that this compound shows magnetodielectric coupling at around 100 K. In addition, unusual behaviour is observed in two polarization loop at room temperature and liquid nitrogen temperature, where coercive field at liquid nitrogen temperature is larger than room temperature. Further, $P$-$E$ loop at liquid nitrogen temperature with different frequencies also affirm that the coercive field and remnant polarization are firstly reduced (but very small value) but when frequency is further increased to 15 Hz and 100 Hz, both of them are enhanced. Therefore, a transition is observed at around 15 Hz in frequency dependent $P_r$ and $E_C$ curve, which may be usually attributed to the generalized pinning and depinning of the dislocation arrays to polarizations.
\end{abstract}
\maketitle

\section{Introduction}
Ferroelectricity and magnetism are very essential in current technology, and the exploration for multiferroic materials, where these two properties are intimately coupled, is of great technological and fundamental importance~\cite{Cheong,Tokura,Spaldin}. But it is very difficult to find the ferroelectric and magnetic properties in a single phase within one material~\cite{Cheong,Nicola-Hill,Khomskii}. Most of the ferroelectrics are transition metal oxides, in which transition metal ions contain empty $d$ orbitals. These positively charged transition metal ions form a molecule with the neighbouring negative oxygen ions. During the formation of the molecule, the collective shift of cations and anions inside a periodic crystal induces bulk electric polarization. The mechanism of the covalent bonding (electronic pairing) in such molecules is the virtual hopping of electrons from the filled oxygen orbital to the empty $d$ orbital of a transition metal ion~\cite{Cheong}. On the contrary, for magnetism partially filled $d$ orbitals of transition metal ions are required, because the spins of electrons occupying completely filled orbitals give zero moment and do not participate in magnetic ordering. The exchange interaction between uncompensated spins of different ions, giving rise to long range magnetic ordering, also results from the virtual hopping of electrons between the ions. In this respect the two mechanisms are not so dissimilar, but the difference in filling of the $d$ shells required for ferroelectricity and magnetism makes these two ordered states mutually exclusive~\cite{Cheong}.
\par
However, still, there is a sizable number of systems which accommodate magnetic cations and consequent possibility of multiferroicity draws much enhanced attention, such as, BiMnO$_3$, BiFeO$_3$,  Pb$_2$MnWO$_6$, Bi$_2$WO$_6$, BiMn$_2$O$_5$etc)~\cite{Seshadri, Volkova, Alonso, Mathieu, Stolen, Shukla}. But based on the microscopic origin of ferroelectricity, multiferroic materials are classified into two different groups, namely, proper (the origin of ferroelectric and magnetism are different) and improper (ferroelectricty originates due to some special kind of magnetic structure) multiferroic materials. In case of proper multiferroic, there are several mechanisms for developing ferroelectricity, such as (i) ferroelectricity due to lone pair (BiMnO$_3$, BiFeO$_3$, PbTiO$_3$, SnTiO$_3$ etc.)~\cite{ Volkova, Seshadri, Nakhmanson, Jin, Hill}, (ii) ferroelectricity due to charge ordering (Pr$_{0.5}$Ca$_{0.5}$MnO$_3$, RMnO$_3$, TbMn$_2$O$_5$, LuFe$_2$O$_4$)~\cite{Khomskii2, Mostovoy, Hur, Ikeda}, (iii) geometric ferroelectricity (YmnO$_3$)~\cite{Spaldin2} etc. On the other hand, in case of improper multiferroic, both noncollinear (TbMnO$_3$, Ni$_3$V$_2$O$_6$, MnWO$_4$)~\cite{Khomskii3, Balatsky, Mostovoy2} and collinear (Ca$_3$CoMnO$_6$)~\cite{Choi} magnetic structures can generate ferroelectricity in the system. Spiral magnetic order, resulting from magnetic frustration, develop exchange striction which is associated with the antisymmetric part of the exchange coupling and constitutes Dzyaloshinsky-Moriya (DM) interaction~\cite{Balatsky, Mostovoy2, Dagotto, Wohlman}. In case of collinear magnetic order, lattice relaxation through exchange striction is associated with the symmetric superexchange coupling~\cite{Choi, Cheong2}.
\par
Surprisingly, short-range magnetic correlation driven multifferoicity has been less explored till date. There are few compounds such as Ca$_3$Co$_2$O$_6$, Er$_2$BaNiO$_5$, etc., which show displacive-type ferroelectricity involving off-centering of the magnetic ion due to short-range magnetic correlation~\cite{Sampathkumaran1, Sampathkumaran2}. Another one langasite compound Pb$_3$TeMn$_3$P$_2$O$_{14}$ (our earlier work) which shows covalency driven modulation of paramagnetism and development of ferroelectricity~\cite{Rafikul-PRB}. In this case, a clear ferroelectric transition is developed near 310 K. Stronger Mn-O covalency redistributes the charges among the other cation-oxygen bonds, which in turn induces crucial electronic changes and consequently, the individual elemental magnetic moment gets changed in the system. As a result, this system shows a first order structural phase transition from $P$321 to $P$3 symmetry at 310 K, which further affects the distribution of moments on the involved atoms locally and as a result a finite magnetoelectric coupling is observed near room temperature~\cite{Rafikul-PRB}. But in our earlier work, we did not discuss about low temperature ($\sim$100 K) phase.
\par
In this paper, we have discussed only the low temperature phase. The temperature dependent XRD and dielectric data (near 100 K) is indicative of further structural disruption locally with decreasing temperature within trigonal symmetric space group of PTMPO compound.  In-depth magnetization, heatcapacity, dielectric, and magnetodielectric studies clearly reveal that short range magnetic correlations are responsible for the dielectric anomaly at around 100 K and consequently shows a magnetodielectric coupling at that temperature.

\section{Methodology}
\subsection{Experimental techniques}
Polycrystalline Pb$_3$TeMn$_3$P$_2$O$_{14}$ has been synthesized by conventional solid state reaction techniques. Pb$_3$TeMn$_3$P$_2$O$_{14}$ (PTMPO) polycrystalline powders were prepared by taking stoichiometric amounts of high purity PbO (Sigma-Aldrich 99.999 \%), TeO$_2$ (Sigma-Aldrich 99.9995 \%), MnO (Sigma-Aldrich 99.99 \%) and NH$_4$H$_2$PO$_4$ (Sigma-Aldrich 99.999 \%). At first, precursors were thoroughly mixed in an agate mortar with ethanol. The mixtures were calcined at 600$^{\circ}$~C for 12 hours in air and finally sintered at 825$^{\circ}$~C for 12 hours in air.
\par
The phase purity and structural characterization of the samples were confirmed by powder X-ray diffraction (XRD) in a Bruker AXS: D8 Advanced x-ray diffarctometer equipped with Cu $K$$_{\alpha}$ radiation. Temperature dependent XRD were carried out at RIGAKU Smartlab (9KW) XG equipped with Cu $K_{\alpha}$ to realize the presence of temperatutre dependent structural phase transitions. The XRD data were analyzed using Rietveld technique and refinements of the crystal structure were done by FULLPROF program~\cite{Carvajal}. The $dc$ Magnetic measurements were carried out using a superconducting quantum interference device (SQUID) magnetometer (Quantum Design, USA) over a temperature range of 2-300 K in magnetic fields upto $\pm$ 5 Tesla and also in a vibrating sample magnetometer (Quantum Design, USA) over a temperature range of 2-400 K. The heat capacity was measured by relaxation method in a quantum design  physical property measurement system (PPMS). The permitivities (for both compounds) were measured using a Hewlett-PacPrecision LCR meter (HP4284A) at an $ac$ level of 1 V mm$^{-1}$. A cryogenic temperature system (Niki Glass LTS-250-TL-4W) was used to control the temperature within the range of 4 - 450 K. The dielectric hysteresis loops were measured using a ferroelectric measurement system (Toyo Corporation FCE-3) equipped with an Iwatsu ST-3541 capacitive displacement meter having a linearity of 0.1\% and a resolution of 0.3 nm. Magnetic field dependent permitivities (at zero Tesla and 9 Tesla) and magnetodielectric measurement were carried out using Keysight Precision LCR meter 4284A in Quantum Design PPMS.

\section {Results and Discussions}
Room temperature X-ray diffraction (XRD) patterns of Pb$_3$TeMn$_3$P$_2$O$_{14}$ (PTMPO) compound has been fitted using a noncentrosymmetric and polar space group $P$3, which is consistent with previous literature reports~\cite{Rafikul-PRB,Silverstein3,Krizan}. Temperature dependent X-ray diffractions have been carried out over a wide temperature range of 5-300 K. All collected XRD patterns have been fitted considering the same trigonal space group $P$3. Reitveld refined XRD at 296 K and 5 K are shown in Figs. 1(a) and (b). Thermal variations of refined lattice parameters, $a$, $b$, $c$ and unit cell volume, are shown in Fig. 1 (c)-(e). An anomaly at $\sim$~120~K has been observed in the temperature dependent lattice parameters as well as unit cell volume variations which signifies structural disruption locally with decreasing temperature within trigonal symmetric space group of PTMPO compound. However, due to only minor changes in position coordinates across the phase transition, the present Rietveld refinements could not conclusively determine the space group of high temperature phase, which is not very unusual in presence of weak lattice distortion~\cite{Sampathkumaran1,Caignaert,Rawat}.

\begin{figure}
\resizebox{8.6cm}{!}
{\includegraphics[70pt,376pt][490pt,707pt]{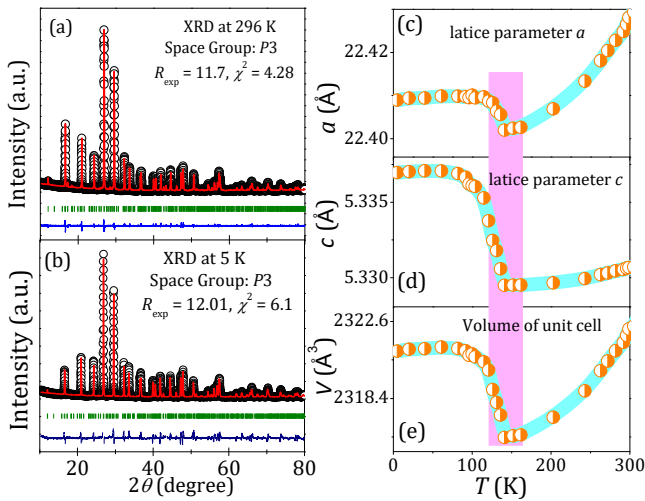}}
\caption{(a)and (b) Rietveld refined XRD at 296 K and 5 K respectively. Black open circles represent the experimental data and red line represents the calculated pattern. The blue line represents the difference between the observed and calculated pattern and green lines signify the position of Bragg peaks. (c), (d) and (e) Thermal variation of lattice parameters and volume of unit cell of PTMPO.}
\end{figure}

\par
Next we focus on the magnetic properties of this compound. Zero-field cooled (ZFC), field cooled cooling (FCC) and field cool heating (FCH) magnetization at 100 Oe were measured in the temperature range 2-300 K, as shown in Fig. 2 (a). Clear antiferromagnetic transitions ($T_N$) around 7 K is observed in both the ZFC, FCC and FCH curve, consistent with previous study~\cite{Rafikul-PRB, Silverstein3, Krizan}. One interesting feature is observed when we have plotted thermal variation of inverse susceptibility data. A significant slope change is observed  above the antiferromagnetic transition (at around 90 K) in the thermal variation of inverse susceptibility data (see Fig. 2(b)), which is further supported by the respective first order derivative curves (see inset of Fig. 2(b)), signifying the short range magnetic correlation at that temperature.

\par
The magnetic transitions are further characterized by the zero field heat capacity measurements ($C_p$ versus $T$), as shown in Fig. 2(c). Lattice part in heat capacity data is determined after fitting the $C_p$ data at high temperature region (100 K-300 K) with Debye-Einstein model~\cite{Debye-Einstein}, which yields a Debye temperature $\Theta_D$ = 314 K and an Einstein temperature $\Theta_E$ = 959 K, and extrapolating to low temperature (down to 2 K, shown in Fig. 2(c) (red line)). The magnetic contribution to the heat capacity is obtained by subtracting the lattice component from the total heat capacity data. A sharp $\lambda$ like anomaly, authentication mark of thermodynamic phase transition into a long range magnetic ordering has been observed near 7 K in the $C_p$ versus $T$ data in Fig. 2(c) and it is in agreement with magnetic susceptibility data. However, a clear hump like behaviour starts just below 120 K in the $C_{mag}$ versus $T$ curve, shown in the Fig. 2(d), which indicates the presence of short range magnetic correlations~\cite{Anupam} and supports further structural disruption locally with decreasing temperature within higher symmetric space group of PTMPO compound.
\begin{figure}[t]
\resizebox{8cm}{!}
{\includegraphics[80pt,437pt][477pt,772pt]{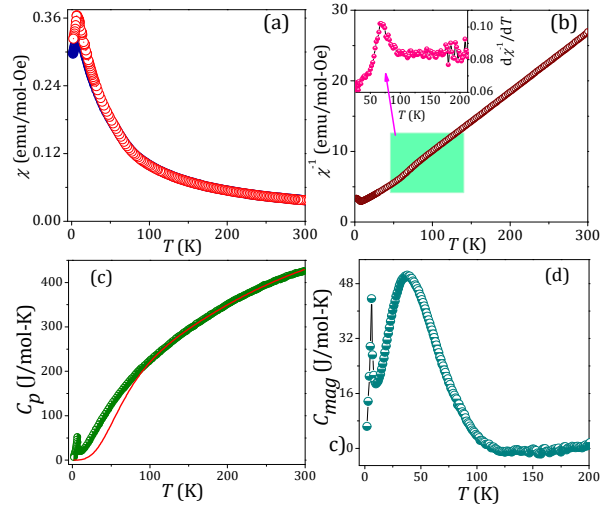}}
\caption{(a) ZFC and FC at 100 Oe of PTMPO and insets show Curie-Weiss fitting of both compounds; Blue solid sphere signifies ZFC and red open circle represents FC curve. (b) Inverse susceptibility data at 100 Oe; Green shaded region indicates the slope change; Inset shows the first order derivative of $\chi (T)$. (c) Heat capacity data (dark yellow) and Debye-Einstein fitting (red line); (d) Magnetic part of heat capacity of PTMPO.}
\end{figure}
\par
The temperature dependence of the real part of the dielectric constant ($\varepsilon'$/$\varepsilon_0$) and dielectric loss ($\tan\delta$)  at different frequencies (1 kHz to 1 MHz) in the temperature range 5 to 400 K have been performed, as shown in the Fig. 3(a). A glass like phase transition (frequency dependent) near 100 K and another frequency independent anomaly  near 310 K appears in the temperature dependent dielectric constant and loss ($\tan\delta$) data. The anomalies near 310 K for PTMPO appear due to the structural phase transition from high temperature nonpolar phase $P$321 to low temperature polar phase $P$3 and this a ferroelectric transition, which has been explained in details in our earlier paper~\cite{Rafikul-PRB}. In this paper we will concern about low temperature glassy phase. This glass like phase transition only for PTMPO comes due to further symmetry breaking with cooling. Further the activation energy ($E_{\text{a}}$ = 132.5 meV) for these relaxations has been calculated from the $\ln$ $f$ versus 1000/$T$ curve using Arrhenius law $f$ = $f_0$$\exp$(-$E_a$/$k_B$$T$) where $E_a$ is the activation energy, $f_0$ is the pre-exponential factor and $k_B$ is the Boltzmann constant (See Figs. 3(b)-(c)).
\begin{figure}
\resizebox{8cm}{!}
{\includegraphics[60pt,276pt][483pt,763pt]{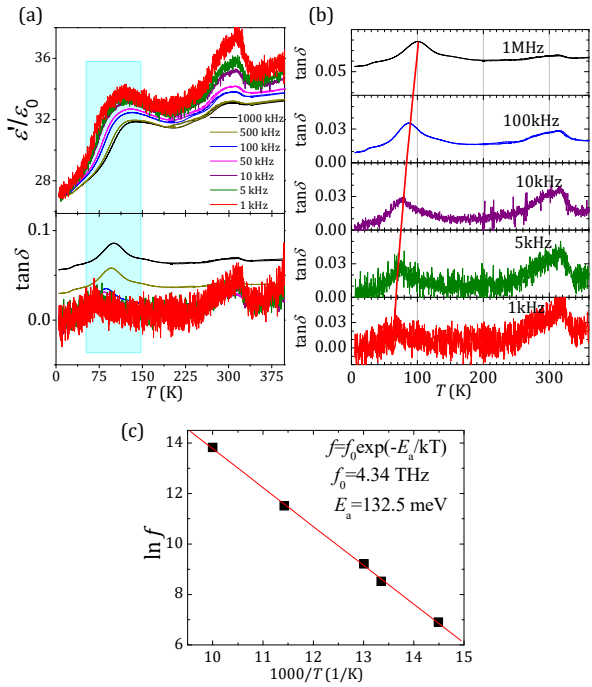}}
\caption{(a) Temperature dependence of real part of dielectric constant $\varepsilon'$/$\varepsilon_0$ and $\tan\delta$ loss data of PTMPO at different frequencies; (b) Frequency dependent $\tan\delta$ loss data; (c) the fitting of the activation energy corresponding to the dielectric loss.}
\end{figure}
\par
As dielectric anomaly, structural phase transition and a short range magnetic correlation are observed at the same temperature (at around 100 K), therefore we study the temperature dependent dielectric behaviour applying zero field and 9 Tesla field. Significant changes at around 100 K and 270 K are observed in the thermal variation of difference between 9 T and zero field capacitance data, as indicated in Fig. 4(a), signifying the presence of magnetodielectric coupling in the system. Explanation of 270 K anomaly has been discussed in our earlier paper~\cite{Rafikul-PRB}. But now we focus on the low temperature anomaly. This magnetodielectric coupling is further established from the isothermal magnetodielectric data at 90 K, as shown in the Fig. 4(b).
\begin{figure}
\resizebox{8.6cm}{!}
{\includegraphics[44pt,444pt][533pt,639pt]{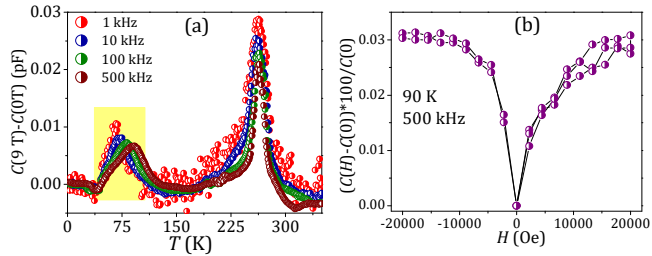}}
\caption{(a) Difference between the thermal variation of capacitance at applied magnetic field 9 T and zero field. (b) Magnetodielectric data at 90 K}
\end{figure}

\begin{figure}[t]
\resizebox{8.6cm}{!}
{\includegraphics[142pt,436pt][512pt,697pt]{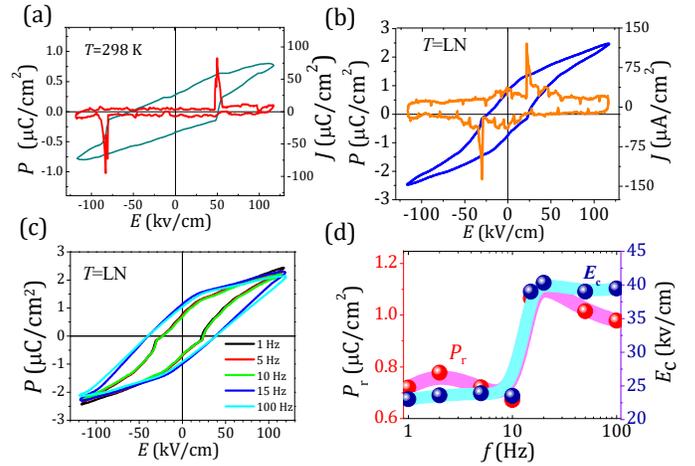}}
\caption{(a) and (b) Electric field variation polarization ($P$) (blue) and switching current density ($J$) (orange) of PTMPO at 298 K and liquid nitrogen temperature respectively; (c) $P$-$E$ loop at different frequencies at liquid nitrogen temperature; (d) Variation of remnant polarization and coercivity with frequency.}
\end{figure}
\par
\par
Further, electric field ($E$) variation polarization ($P$) have been carried out at room temperature and liquid nitrogen temperature, as shown in Fig. 5(a) and (b) respectively. Clear $P-E$ loop with saturation and remnant polarization 0.83 $\mu$C/$cm^2$ and 0.3 $\mu$C/$cm^2$ respectively, is observed at 300 K. Square loop with switching current peaks at room temperature in the switching current density ($J$) vs electric field ($E$) curve, indicates the ferroelectricity of the sample at room temperature, shown in Fig. 5(a). There are two reasons behind the polarization of PTMPO at room temperature. One is geometrically polar structure and second is the  stereochemical active lone pair of Pb$^{2+}$ ($ns^2$). Surprisingly, careful analysis of two polarization loop says that coercive field at liquid nitrogen temperature is larger than room temperature, which is abnormal. Further, $P$-$E$ loop at liquid nitrogen temperature with different frequencies has been taken, as shown in Fig. 5(c). In this case we get another abnormality, as frequency is increased (from 1 Hz to 10 Hz), the coercive field and remnant polarization are firstly reduced (but very small value) but when frequency is further increased to 15 Hz and 100 Hz, both of them are enhanced. $P_r$ and $E_C$ versus frequency ($f$) have been plotted in the Fig. 5(d), while we can clearly observe a transition around 15 Hz. The kind of frequency-dependent properties are usually attributed to the generalized pinning and depinning of the dislocation arrays to polarizations~\cite{Dislocation}.

\section {Conclusion}
Here we have reported the results of structural, magnetic and dielectric properties in details. An anomaly at $\sim$~120~K in the temperature dependent lattice parameters as well as unit cell volume variations obtained from Rietveld refinement of temperature dependent X-ray diffractions, signifies structural disruption locally with decreasing temperature within trigonal symmetric space group of PTMPO compound. A significant slope change in the inverse susceptibility data and hump like behaviour in heat capacity data clearly affirm the presence of short range magnetic correlation above the antiferromagnetic transition temperature. A frequency dependent anomaly in thermal variation of dielectric constant is also observed at the same temperature range of short range magnetic correlation. A very small but significant changes in between the with field ($H$ = 9T) and without field data near $\sim$100 K as well as magnetodielectric measurement signify the presence of magnetoelectric coupling in the system. As a result, we may conclusively state that the structural phase transition at around 120 K affects the distribution of moments on the involved atoms locally and as a result a finite magnetoelectric coupling is observed near about 100 K. Another interesting thing is observed in the frequency dependent $P_r$ and $E_C$ curve, which are usually ascribed to the generalized pinning and depinning of the dislocation arrays to polarizations.

\section {Acknowledgment}
R.A.S thanks CSIR, India and IACS for a fellowship. S.R thanks DST, India for funding (project no.CRG/2019/003522), Technical Research Center of IACS, Indo-Italian POC for support to carry out experiments in Elettra, Italy and Laboratory for Materials and Structures, Collaborative Research Projects for providing experimental facilities.


\begin{thebibliography}{99}
\bibitem{Cheong} S.-W. Cheong, and M. Mostovoy, Nature matreials {\bf6}, 13 (2007).
\bibitem{Tokura} T. Kimura, T. Goto, H. Shintani, K. Ishizaka, T. Arima, and Y. Tokura, Nature (London) {\bf426}, 55 (2003).
\bibitem{Spaldin} N. A. Spaldin and M. Fiebig, Science {\bf309}, 391 (2005).
\bibitem{Nicola-Hill} N. A. Hill, J. Phys. Chem. B ,{\bf104}, 6694$-$6709 (2000).
\bibitem{Khomskii} D. Khomskii, Bull. Am. Phys. Soc. C 21.002 (2001).
\bibitem{Seshadri} R. Seshadri and N. A. Hill, Chem. Mater {\bf13}, 2892-2899 (2001).
\bibitem{Volkova} L. M. Volkova and D. V. Marinin, J Supercond. Nov. Magn. {\bf24}, 2161-2177 (2011).
\bibitem{Alonso} S. A. Larregola, J.A. Alonso, M. Alguero, R. Jimenez, E. Suard, F. Porcher and J.C. Pedregosa, Dalton Trans., {\bf39} 5159-5165 (2010).
\bibitem{Mathieu} S. A. Ivanov, A. A. Bush, A. I. Stash, K. E. Kamentsev, V. Ya. Shkuratov, Y. O. Kvashnin, C. Autieri, I. D. Marco, B. Sanyal, O. Eriksson, P. Nordblad, and  R.  Mathieu, Inorg. Chem. {\bf55}, 2791$-$2805 (2016).
\bibitem{Stolen} C. E. Mohn, and S. St${\o}$len, Phys. Rev. B {\bf83}, 014103 (2011).
\bibitem{Shukla} D. K. Shukla, S. Mollah, Ravi Kumar, P. Thakur, K. H. Chae, W. K. Choi, and A. Banerjee, J. Appl. Phys. {\bf104}, 033707 (2008).

\bibitem{Jin} X. He and K. Jin, Phys. Rev. B {\bf94}, 224107 (2016).
\bibitem{Hill} N. A. Hill and K. M. Rabe, Phys. Rev. B {\bf59}, 8759 (1999).
\bibitem{Nakhmanson} K. C. Pitike, W. D. Parker, L. Louis and S. M. Nakhmanson, Phys. Rev. B {\bf91}, 035112 (2013).
\bibitem{Khomskii2} D. V. Efremov, J. van den Brink, and D. I. Khomskii, Nature Mater. {\bf3}, 853 (2004).
\bibitem{Mostovoy} S. W. Cheong and M. V. Mostovoy, Nature Mater. {\bf6}, 13 (2007).
\bibitem{Hur} N. Hur, S. Park, P. A. Sharma, J. S. Ahn, S. Guha and S-W. Cheong, Nature {\bf429}, 392 (2004).
\bibitem{Ikeda} N. Ikeda, H. Ohsumi, K. Ohwada, K. Ishii, T. Inami, K. Kakurai,Y. Murakami, K. Yoshii, S. Mori, Y. Horibe and H. Kito, Nature
{\bf436}, 1136 (2005).
\bibitem{Spaldin2} B. B. van Aken, T. T. M. Palstra, A. Filippetti and N. A. Spaldin, Nature Mater. {\bf3}, 164 (2004).
\bibitem{Khomskii3} D. Khomskii Physics {\bf2}, 20 (2009).
\bibitem{Balatsky} H. Katsura, N. Nagaosa and A. V. Balatsky, Phys. Rev. Lett. {\bf95}, 057205 (2005).
\bibitem{Mostovoy2} M. V. Mostovoy, Phys. Rev. Lett. {\bf96}, 067601 (2006).
\bibitem{Choi} Y. J. Choi, H. T. Yi, S. Lee, Q. Huang, V. Kiryukhin and S.-W. Cheong, Phys. Rev. Lett. {\bf100}, 047601 (2008).
\bibitem{Dagotto} I. A. Sergienko and E. Dagotto, Phys. Rev. B {\bf73}, 094434 (2006).
\bibitem{Wohlman} A. B. Harris, T. Yildirim, A. Aharony and O. Entin-Wohlman, Phys. Rev. B {\bf73}, 184433 (2006).
\bibitem{Cheong2} L. C. Chapon, P. G. Radaelli, G. R. Blake, S. Park, and S.-W. Cheong, Phys. Rev. Lett. {\bf96}, 097601 (2006).
\bibitem{Sampathkumaran1} T. Basu, V. V. R. Kishore, S. Gohil, K. Singh, N. Mohapatra, S. Bhattacharjee, B. Gonde, N. P. Lalla, P. Mahadevan, S. Ghosh and E. V. Sampathkumaran, Sci. Rep. {\bf4}, 5636 (2014).
\bibitem{Sampathkumaran2} T. Basu, K. K. Iyer, K. Singh and E. V. Sampathkumaran, Sci. Rep. {\bf3}, 3104 (2013).
\bibitem{Rafikul-PRB} R. A. Saha, A. Halder, T. Saha-Dasgupta, D. Fu, M. Itoh, and S. Ray, Phys. Rev. B {\bf101}, 180406(R) (2020).
\bibitem{Carvajal}  J. Rodriguez Carvajal, Physica B {\bf192}, 55 (1993).
\bibitem{Silverstein3} H. J. Silverstein, A. Huq, M. Lee, E. S. Choi, H. Zhou and C. R. Wiebe, J. Solid State Chem. {\bf221}, 216$–$223 (2015).
\bibitem{Krizan} J.W. Krizan, C. de la Cruz, N.H. Andersen and R.J. Cava, J. Solid State Chem. {\bf 203}, 310$-$320(2013).
\bibitem{Caignaert} V. Caignaert, A. Maignan, K. Singh, Ch. Simon, V. Pralong, B. Raveau, J. F. Mitchell, H. Zheng, A. Huq, and L. C. Chapon, Phys. Rev. B {\bf88}, 174403 (2013).
\bibitem{Rawat} S. Goswami, P. D. Babu, and R. Rawat, J. Phys.: Condens. Matter {\bf31}, 445801 (2019).
\bibitem{Debye-Einstein} C. A. Martin, J. Phys.: Condens. Matter {\bf3}, 5967 (1991).
\bibitem{Anupam} J. Sannigrahi, J. Sichelschmidt, B. Koo, A. Banerjee, S. Majumdar and S. Kanungo, J. Phys.: Condens. Matter {\bf31}, 245802 (2019).
\bibitem{Dislocation} H. H. Wu, S. G. Cao, J. M. Zhu, T. Y. Zhang, Acta Mech. {\bf228}, 2811$-$2817 (2017).
\end{thebibliography}
\end{document}